\documentclass[prl,twocolumn,showpacs,floatfix]{revtex4}

\usepackage[latin1]{inputenc}
\usepackage[T1]{fontenc}
\usepackage[english]{babel}
\usepackage{times}
\usepackage{graphicx}
\usepackage{hyperref}
\hypersetup{%
	pdfauthor   = {G. Lang, H.-J. Grafe, D. Paar, F. Hammerath, K. Manthey, G. Behr, J. Werner, and B. Buechner},%
	pdftitle    = {Nanoscale electronic order in iron pnictides}%
	}

\usepackage{fancyhdr}
\def\OFx{O$_{1-x}$F$_x$}

\begin{document}

\title{Nanoscale electronic order in iron pnictides}
\author{G. Lang,\textsuperscript{1,*} H.-J. Grafe,\textsuperscript{1} D. Paar,\textsuperscript{1,2} F. Hammerath,\textsuperscript{1} K. Manthey,\textsuperscript{1} G. Behr,\textsuperscript{1} J. Werner,\textsuperscript{1} and B. B\"uchner\textsuperscript{1}}
\affiliation{
\textsuperscript{1}IFW Dresden, Institute for Solid State Research, P.O. Box 270116, D-01171 Dresden, Germany \\
\textsuperscript{2}Department of Physics, Faculty of Science, University of Zagreb, P.O. Box 331, HR-10002 Zagreb, Croatia}

\begin{abstract}
The charge distribution in {\it R}FeAs\OFx\ ({\it R}=La, Sm) iron pnictides is probed using As nuclear quadrupole resonance.
Whereas undoped and optimally-doped or overdoped compounds feature a single charge environment, two charge environments are detected in the underdoped region. Spin-lattice relaxation measurements show their coexistence at the nanoscale. Together with the quantitative variations of the spectra with doping, they point to a local electronic order in the iron layers, where low- and high-doping-like regions would coexist. Implications for the interplay of static magnetism and superconductivity are discussed.
\end{abstract}


\maketitle

Strong electronic correlations, notably present in several transition metal oxides, give rise to a broad range of exotic electronic states, in particular high-temperature superconductivity.
Major interest has followed from the recent discovery \cite{Kamihara2008} of the latter in iron pnictides, with critical temperatures second only to those of copper-based materials.
Iron pnictides feature electronically active FeAs layers with a	multiorbital character, separated by {\it R}(\OFx) layers ({\it R}=rare earth) for the ``1111'' family studied here.
On doping with electrons through fluorine substitution, the antiferromagnetic order of the parent material \mbox{{\it R}FeAsO} is suppressed and a superconductivity region shows up in the phase diagram. While the itinerant nature of the undoped compound contrasts with the Mott insulating character of the undoped cuprates, the renewed proximity of static magnetism and superconductivity is fueling the idea that competition with another ground state is crucial to high-temperature superconductivity.
However, reports in underdoped pnictides are so far conflicting, with microscopic \cite{Drew2009,Laplace2009,Julien2009} or mesoscopic \cite{Park2009,Julien2009} coexistence, a second-order boundary \cite{Zhao2008}, or a first-order boundary \cite{Luetkens2009}.
While this can be accounted for to some extent by differing sensitivities of the experimental probes, in particular to disordered magnetism, it remains unclear whether intrinsic electronic inhomogeneities and an associated order, short-range or more, can show up such as in cuprates \cite{Tranquada1995,Pan2001,Kohsaka2007}, nickelates \cite{Chen1993,Wochner1998} and manganates \cite{Brink1999,Weber2009}.
In the presence of such inhomogeneities, both the magnetic ordering and superconductivity ground states may be altered, impacting their interplay or competition.

Any intrinsic inhomogeneities may be sensitive to the presence of impurities within the FeAs planes, as in Co-doped BaFe$_2$As$_2$, or to the electrostatic potential associated to the dopant layers. As the 1111 pnictides correspond to the case where the dopants are the farthest away from the FeAs layers and with rare earth layers in between, they represent the situation of minimal influence of the dopant layers on the intrinsic electronic properties. Therefore, we investigated the charge distribution in {\it R}FeAs(\OFx) ({\it R}=La, Sm) 1111 compounds, over the 0$\le$$x$$\le$0.15 doping range, using nuclear quadrupole resonance (NQR) measurements. Our study shows the presence of an electronic inhomogeneity in the underdoped region of the phase diagram, ascribed to a nanoscale electronic order and independent from the low-temperature behavior.
Revealing that the phase diagram has to be understood beyond a homogeneous picture with ground state competition, this result also supports the idea that local electronic order is a common feature of strongly-correlated systems, notably in transition metal oxides \cite{Dagotto2005a}.

All NQR measurements were carried on powder samples, which were prepared and studied using x-ray, susceptibility, resistivity, and $\mu$SR measurements as described previously \cite{Luetkens2009,Kondrat2009,Hess2009,Klingeler2010}.
Undoped La and undoped/4\% ($x$=0.04) Sm samples were shown to display magnetic ordering at $T_M$=138~K (La) and $T_M$=137/95~K (Sm), with the accompanying structural transition at $T_S$=156~K (La) and $T_S$=160/140~K (Sm). The 5\%/7.5\%/10\%/15\% (La) and 6\%/8\%/10\% (Sm) samples display bulk superconductivity below $T_c$=20/21/26/11~K (La) and $T_c$=36/45/52~K (Sm).
To probe the atomic populations corresponding to potential electronic inhomogeneities, NQR takes advantage from the fact that a nucleus with a nuclear spin $I$$>$$1/2$ features a finite electric quadrupole moment.
In the presence of a finite electric field gradient (EFG) at the nuclear site, the degeneracy of the corresponding nuclear energy levels is lifted, and probing by radiofrequency irradiation can be performed.
As the EFG stems from the surrounding charge distribution, peculiarities of the latter can be inferred from the determination of the EFG histogram in the sample.
Since $I$$=$$1/2$ for iron, the $^{75}$As nuclei ($I$$=$$3/2$) were used as NQR probes. One measures then the quadrupole frequency $\nu_Q \propto V_{zz} \sqrt{1+\eta^2/3}$, where $V_{zz}$ and $\eta$ are respectively the highest eigenvalue and the asymmetry of the EFG tensor.
The proximity of the As ions to the iron layers helps to retain high sensitivity to electronic changes, which may be further helped by their large polarizability \cite{Berciu2009}.

\begin{figure}[pbth]
\includegraphics[width=80mm]{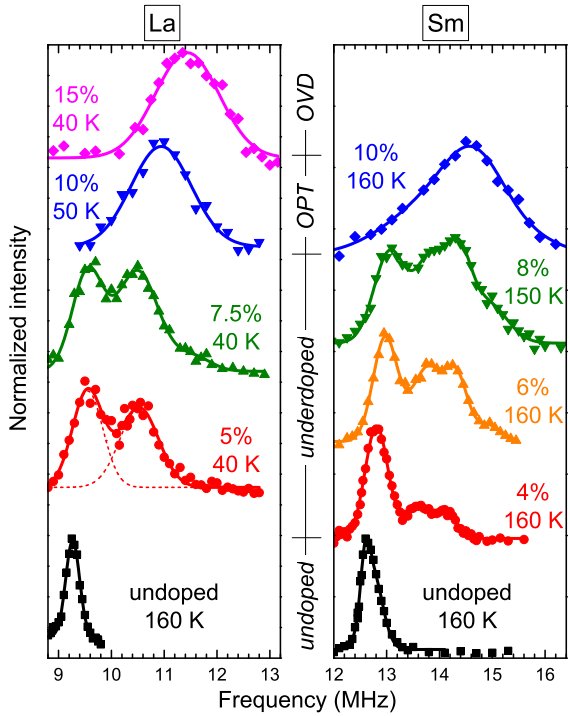}
\caption{(Color online) $^{75}$As NQR spectra of RFeAs(\OFx) (R=La, Sm). ``OPT'' and ``OVD'' refer to optimally doped and overdoped samples. Fits including up to three (La) or four (Sm) Gaussians are shown as full lines, with the two-Gaussian fit for $x$=0.05 (La) detailed as an example.}
\label{fig:spectra}
\end{figure}

\begin{figure}[pbth]
\includegraphics[width=75mm]{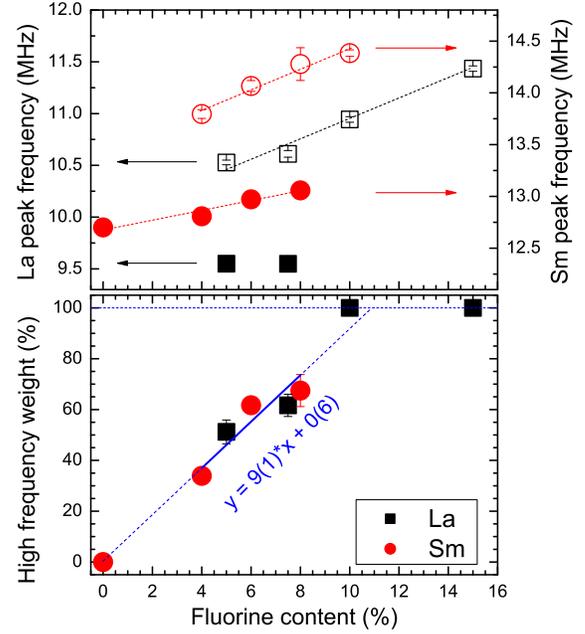}
\caption{(Color online) (Upper panel) Doping dependence of the spectral peak frequencies, taken at the centers of gravity of the low- and high-frequency peaks of Fig.~\ref{fig:spectra}. For each rare earth, only points taken at similar temperatures are shown. Dotted lines are visual guides. (Lower panel) Doping dependence of the relative weight of the high-frequency peak. The solid blue line is a linear fit for 0.04$\le$$x$$\le$0.08. Dotted lines are visual guides.}
\label{fig:specfit}
\end{figure}

We start by examining the static properties, i.e., the histogram of the electric quadrupole frequency $\nu_Q$.
A standard $\frac{\pi}{2}$--$\tau$--$\pi$ spin echo sequence was used, with $\tau$=24--34~$\mu$s and integration of the full echo. The $\tau$ values, as well as the pulse repetition rate, were checked to be small enough that spin-spin or spin-lattice relaxation contrast is limited to at most a few percent of the relative line intensities.
Spectra were corrected for frequency-dependence of the intensity.
On doping (see Fig.~\ref{fig:spectra}), the frequency distribution shifts to higher values, in keeping with previous observations \cite{Grafe2009,Mukuda2008} and in disagreement with local-density approximation calculations \cite{Grafe2009,Jeglic2009}.
In the undoped limit, the single narrow line agrees with a single well-defined charge environment for all As nuclei.
The line is broadened in the optimally-doped or overdoped limit (La 10\%/15\%, Sm 10\%), likely reflecting the structural disorder of the fluorine dopants and some moderate fluorine concentration inhomogeneities.
In the underdoped region (La 5\%/7.5\%, Sm 4\%/6\%/8\%), two fairly broad peaks are observed, with further structuring of the high-frequency peak for Sm samples.
Spectral weight is transferred directly to high frequency on doping, with a roughly linear dependence extrapolating to the optimal doping (see lower panel of Fig.~\ref{fig:specfit}).
Crucial questions are then the difference in nature and the scale of coexistence of the two corresponding sets of charge environments.

\begin{figure}[pbth]
\includegraphics[width=75mm]{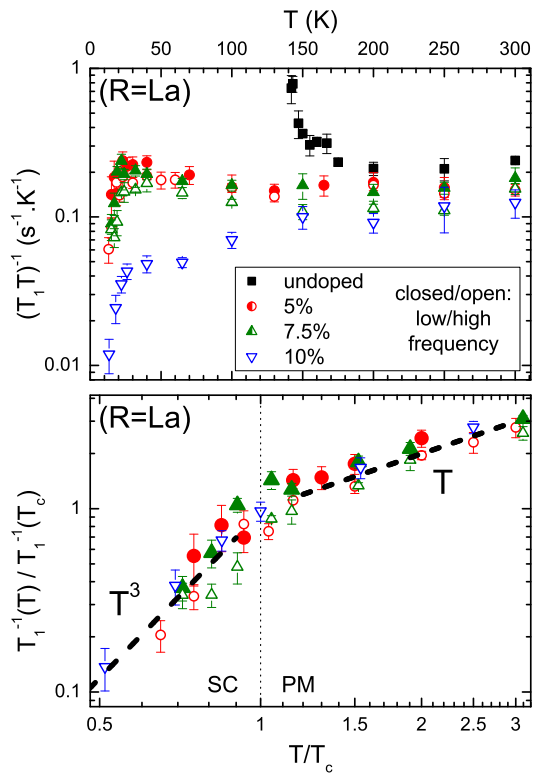}
\caption{(Color online) $^{75}$As spin-lattice relaxation in LaFeAs(\OFx). (Upper panel) Temperature dependence of $(T_1T)^{-1}$ with $T_1^{-1}$ the spin-lattice relaxation rate as measured on each peak of Fig.~\ref{fig:spectra}, with closed (open) symbols corresponding to low (high) frequencies. The measurements are done at the T-dependent peak frequency (undoped), 9.7 and 10.6~MHz (both 5\% and 7.5\%), and 11~MHz (10\%). It was checked on the 5\% sample that the spectrum is preserved on crossing $T_c$. (Lower panel) Temperature-dependence of $T_1^{-1}$ at low temperature, with horizontal scaling by $T_c$  (as determined from initial characterization) and vertical scaling to obtain coincidence about $T_c$. Symbols for the low-frequency peak in the underdoped samples are oversized for emphasis. Dashed lines are visual guides. ``SC'' and ``PM'' refer to superconductivity and paramagnetism.}
\label{fig:relaxation}
\end{figure}

A direct explanation would be phase separation on a macroscopic or mesoscopic scale, with the difference in peak positions (see upper panel of Fig.~\ref{fig:specfit}) indicating low and high doping regions.
Beyond incompatibility with initial x-ray characterization, this can be tested using $T_1^{-1}$ spin-lattice relaxation rate measurements in the La samples.
For low-energy spin fluctuations over all wave vectors $\mathbf q$,
$$(T_1T)^{-1} \propto \sum_{\mathbf q} \left| ^{75}A_{hf}({\mathbf q}) \right|^2 \frac{\chi''({\mathbf q},\omega_{rf})}{\omega_{rf}},$$
where $^{75}A_{hf}$ is the As hyperfine form factor, $\chi''$ is the dynamic spin susceptibility, and $\omega_{rf}$ is the irradiation frequency.
The relaxation was measured using the inversion recovery sequence, fitting the recovery curve to
$$M(t) = M_0 (1 - f \exp(-(3t/T_1)^\lambda)),$$
where $M_0$ is the magnetization at thermal equilibrium, $f$ accounts for incomplete inversion, and $\lambda$ accounts for spreading of the $T_1$ value. If set free, $\lambda$ tended to drift below unity below $T_c$, with no significant effect on the extracted $T_1$ value.
Figure \ref{fig:relaxation} shows for each spectral peak the temperature dependence of $(T_1T)^{-1}$ (upper panel) and of the same $T_1^{-1}$ data, rescaled about $T_c$ (lower panel).
While $(T_1T)^{-1}$ in the undoped material tends to diverge on approaching the magnetic transition as expected, all other peaks show no signature of magnetic ordering.
They reflect however a superconducting transition at low temperature as seen from the rapid decrease of the relaxation, with a $T_1^{-1}$ behavior broadly consistent with observed power laws \cite{Grafe2008,Nakai2008}.
For the underdoped samples in the paramagnetic state, the two spectral peaks feature similar $(T_1T)^{-1}$ behaviors, very different from the progressive suppression of low-energy excitations observed at optimal doping.
If phase separation would occur even on a rather small scale (several nanometers or more), the volume fraction corresponding to each peak would exhibit fluctuations specific to the then necessarily different doping levels. In light of the spectra, the relaxation contrast should then be much larger, with the high-frequency peak relaxation closer to developing a decrease similar to that at optimal doping. Here, the similar weak Curie-Weiss behavior above $T_c$ and the moderate difference in amplitude show a sharing of electronic properties, i.e., the coexistence of the two charge environments at the nanoscale.
The adequacy of using the same $T_c$ to rescale the $T_1^{-1}$ data for the two peaks at a given doping (lower panel of Fig.~\ref{fig:relaxation}) is consistent with this conclusion, at least for $x$$=$0.075.
Note that a comparable study cannot be performed on the Sm samples as $T_1^{-1}$ is swamped by the contribution of the Sm magnetic moments. However, beyond the similarity to the La spectra, the increase of both peak frequencies on transferring spectral weight (see upper panel of Fig.~\ref{fig:specfit}) is clearly inconsistent with a mere change in the proportion of two well-separated phases, and suggests mutual influence at the local level.

Crucial questions are the origin of this coexistence and the spatial charge arrangement corresponding to it.
A possibility is fluorine control of the two charge environments at the As sites through direct electrostatic effect or local doping of the FeAs layers.
To properly account for the weights of the peaks, each fluorine must influence on average roughly nine As ions (see lower panel of Fig.~\ref{fig:specfit}).
This rules out a one-to-one effect of each dopant on the nearest arsenic, contrary \cite{Singer2002} to the effect of Sr\textsuperscript{2+} on Cu ions in La$_{2-x}$Sr$_x$CuO$_4$.
If a single fluorine would affect a patch of As ions, the linear growth of the high-frequency weight with doping suggests that no sizable overlap of these patches develops as the doping rises.
In particular, the fact that the linear behavior extrapolates well to the optimal doping indicates that in the latter case there would be a perfect arrangement of the patches, i.e., fluorine ordering.
While dopant ordering is known to occur in compounds such as sodium cobaltates \cite{Alloul2009}, there is no experimental hint of it in pnictides.
Therefore, the inhomogeneity likely arises not from the dopants but from an electronic instability in the FeAs layers, due to competing interactions.

The resolved spectral features and their smooth doping dependence indicate rather well-defined and reproducible local electronic environments, static on the slow (microsecond) time scale of NQR.
As seen on the upper panel of Fig.~\ref{fig:specfit}, the low-frequency and high-frequency environments are reasonably well connected, respectively, to the undoped and optimally-doped or overdoped environments. Together with the linear spectral weight transfer extrapolating to the optimal doping, this suggests a local electronic order with a varying ratio of low-doping-like and high-doping-like regions.
Expected to feature larger intrinsic magnetic fluctuations, the low-doping-like regions would account for the relaxation response of the whole system.
Since doping corresponds to changes in the iron $3d$ orbital occupancies, it must, however, be noted that what appears to be a difference in total occupancy (charge ordering) around As sites may also have an orbital character (orbital ordering) \footnote{Note that even small differences in orbital occupancy can result in a large change of quadrupole frequency, due to the fact that the intervening distances are very small and the arsenic electronic cloud is easily distorted.}.
This scenario would be of interest in light of the argued link \cite{Krueger2009,Kubo2009} between orbital ordering and static magnetism.
More generally, the presence of a local electronic order is reminiscent of the situation observed for instance in cuprates \cite{Tranquada1995,Kohsaka2007} and manganates \cite{Brink1999,Weber2009}, such as stripe or checkerboard order, static or dynamic. This supports the widespread presence of electronic inhomogeneities in correlated systems \cite{Dagotto2005a}, even in the presence of a homogeneous ground state.

\begin{figure}[pbth]
\includegraphics[width=80mm]{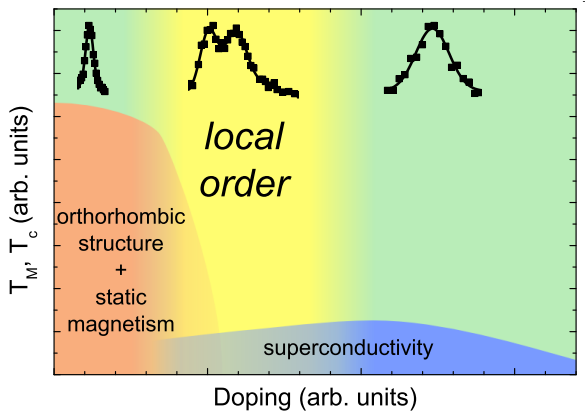}
\caption{(Color online) Electronic phase diagram for the 1111 pnictides. $T_M$/$T_c$ denote the magnetic and superconducting transition temperatures. A local electronic order develops in the underdoped region, with superconductivity being favored over static magnetism. At the top are shown typical NQR spectra, taken from Fig.~\ref{fig:spectra}.}
\label{fig:diagram}
\end{figure}

Being common to all underdoped samples, the local order is no direct explanation for any ground state coexistence.
However, our data show that $T_c$ remains high despite significant magnetic fluctuations, just as the superfluid density is comparatively more depressed \cite{Luetkens2009}, possibly indicating a proximity effect in nonintrinsically superconducting low-doping sample regions.
On the contrary, on doping, the static magnetism tends to be disordered \cite{Drew2009,Sanna2009}, with a large reduction of $T_M$.
Therefore we propose that, beyond direct ground state competition, the electronic inhomogeneity of the underdoped region of the phase diagram influences the transition from static magnetism to superconductivity, as shown schematically on Fig.~\ref{fig:diagram}.
Starting from high doping, superconductivity is unhindered if not helped by the local order and disappears \cite{Luetkens2009,Drew2009,Zhao2008} only at its low-doping end, where the high proportion of low-doping-like regions would allow static magnetism to shoot up, before recovering electronic homogeneity close to the undoped limit.
Note that, as is the case in other systems \cite{Uemura2007}, the local order occurs where a quantum critical point could have been expected in the phase diagram.
A crucial question is then whether the local environments correspond to specific points of the phase diagram, as is seen \cite{Mohottala2006} in La$_{2-x}$Sr$_x$CuO$_{4+y}$ with separation in stripe-ordered and optimally doped superconducting regions.

In conclusion, using As NQR, we show the presence of two charge environments at the nanoscale in underdoped 1111 iron pnictides. These can be ascribed to a local electronic order in the iron layers, in line with electronic inhomogeneities observed in other correlated compounds.
We propose a phase diagram in which this local order serves as the electronic background for the interplay of static magnetism and superconductivity, beyond direct competition.
An intriguing scenario, to be addressed in future work, is the possible orbital nature of this local order, which could suggest that a primary factor in the phase diagram is real-space orbital competition, where sensitivity to structural details would yield seemingly different phase diagrams.
In light of the reported importance of the Fermi surface topology \cite{Korshunov2008a}, this would represent a significant change of perspective on this system.

We acknowledge discussions with H. Alloul, J. Bobroff, J. Hamann-Borrero, C. Hess, V. Kataev, D. Morr, and S. Wurmehl, as well as experimental support from A. K\"ohler, S. Ga\ss\ and R. Vogel. This work was supported by the Deutsche Forschungsgemeinschaft, through FOR 538, SPP 1458, and Grant No. Be1749/12. G.L. acknowledges support from the Alexander von Humboldt Foundation.

\textsuperscript{*}g.m.lang@ifw-dresden.de

\bibliographystyle{unsrt}

\end{document}